\documentclass[published]{JHEP} 

\usepackage{epsfig}

\def\a{\alpha}
\def\b{\beta}

\def\d{\delta}
\def\D{\Delta}
\def\e{\epsilon}

\def\vphi{\varphi}

\def\pa{\partial}

\def\cN{{\cal N}}

\def\shalf{\frac{1}{2}}

\def\w{\wedge}

\def\be{\begin{equation}}
\def\ee{\end{equation}}
\newcommand{\brr}{\begin{eqnarray}}
\newcommand{\err}{\end{eqnarray}}

\title{Brane Worlds, the Cosmological Constant and String Theory}
 
\author{S. P. de Alwis, A. T. Flournoy and N. Irges\\
        Department of Physics, Box 390,
        University of Colorado, Boulder, CO 80309.\\
        E-mail: \email{dealwis@pizero.colorado.edu,
                       flournoy@pizero.colorado.edu,
                       irges@pizero.colorado.edu}}

\received{November 29, 2000}
\accepted{January 22, 2001}
\JHEP{01(2001)027}
 
\preprint{\hepth{0004125}}      

\abstract{
We argue that traditional methods of 
compactification of string theory make it 
very difficult to understand {\it how} the 
cosmological constant becomes zero. String 
inspired models can give zero cosmological 
constant after fine tuning but since string theory has
no free parameters it is not clear that 
this is allowed. Brane world
scenarios on the other hand while they do 
not answer the question as to {\it why} the 
cosmological constant is zero do actually allow
a choice of integration constants that permit 
flat four space solutions. In this paper we 
discuss gauged supergravity realizations of 
such a world. To the extent that this starting 
point  can be considered a low energy effective 
action of string theory (and there is some 
recent evidence supporting this) our model 
may be considered a string theory realization of this scenario.
}

\keywords{D-branes, Superstring Vacua, Supergravity Models.}
 
\begin{document}
 
\section{Introduction}

The traditional method of compactification of superstring theory involves
writing the ten dimensional manifold as a direct product of 
a non-compact Minkowskian four manifold and a 
compact Euclidean six manifold $M_{10}=M_{4}\otimes M_{6}$, 
with the size of the compact space
being set by the string scale. The four manifold is identified with
the space we observe around us and is therefore taken 
to be flat. The compact manifold is taken to be a Ricci 
flat K${\ddot {\rm a}}$hler manifold (i.e. a Calabi-Yau space or an orbifold) 
so that we obtain ${\cal N}=1$ supersymmetry (SUSY) in the four 
manifold \cite{gsw}. The latter is required for at least 
two reasons. The first is the so-called hierarchy problem 
involving the stabilization of the weak scale to quantum 
corrections. With (softly
broken) ${\cal N} =1$ SUSY one can protect this hierarchy. 
The other 
issue is that of the cosmological constant (CC). Again 
supersymmetry offers the hope of stabilizing 
this against quantum corrections i.e. if the 
ultraviolet theory (say at the string scale) 
has zero cosmological constant then as long as 
SUSY is preserved it will remain zero. In this 
case case, however, once  SUSY is broken, it is difficult to see how
one can prevent a CC, at least of order the mass splitting between 
fermions and bosons, from being generated. 

Actually getting a realistic model where this is achieved would 
be progress, since there would also be standard model
phase transitions which are also of the same order 
and could in principle cancel the SUSY breaking effects. 
Nevertheless at this point
it is, we believe, fair to say that even this has 
not yet been achieved in a natural way, since 
typical cosmological constants that are generated 
upon SUSY breaking are at an intermediate scale 
of ${\cal O}(10^{11}GeV)$ (which is the actual scale of 
SUSY breaking in the so-called hidden sector). 

To get a zero (or small) CC requires fine tuning of some 
parameter in the scalar potential of ${\cal N} =1$ 
supergravity. However string theory has no tunable 
parameters. In fact it is not very clear how to even 
generate the potential for the moduli (the dilaton, the size and
shape of the compact manifold) but it is generally agreed that, either 
stringy or low energy field theoretic, non-perturbative effects will
give such a potential and stabilize these moduli.
\footnote{For a recent discussion highlighting the 
problems involved with references to earlier literature see \cite{bda}.}

Let us first think about the field theoretic stabilization mechanism. 
One possibility is to have several 
gauge groups in the hidden sector whose gauginos condense. 
This is the so-called race track model. The effective 
superpotential needs to be  a sum of three exponentials 
of the moduli fields, two are needed to get a weak gauge 
coupling minimum and an intermediate scale of SUSY breaking 
and a third is needed in order to fine tune the cosmological 
constant to zero. This model has various problems associated 
with the steepness of the potential (see \cite{bda} for 
a recent discussion) but even apart from that it seems 
unlikely that the sort of fine tuning that is required 
is allowed by string theory. An alternative would be to 
have the moduli stabilized by string scale 
physics. In this case however one needs to find field 
theoretic models which solve the practical cosmological constant problem i.e.
obtain a model with CC only at the weak scale as 
advocated in \cite{bda}. Even this seems a non-trivial task; 
in particular, it seems that one needs a constant in 
the superpotential to do so and to get a zero (or small) 
cosmological constant  would seem to require a fine 
tuning that would really be at variance with  what one 
expects from string theory. Similar remarks apply to 
all models proposed so far for getting standard model 
physics out of string theory (including many brane 
world type scenarios based on type I string theory) once one tries to get
flat four dimensional space after SUSY breaking.

The main point of this paper is to propose a string 
theoretic brane world scenario for getting a zero 
cosmological constant based on arguments 
sketched in \cite{sda}. This would 
imply a radical departure from what happens in the 
standard string compactification 
models discussed in the above paragraphs. 
Instead of compactifying to
four flat dimensions with ${\cal N}=1$ 
supersymmetry, we compactify
the ten dimensional theory (here we will 
concentrate on type IIB)
on a five sphere (or squashed sphere) to 
five dimensions \cite{Bremer}. 
In contrast to the standard compactifications 
we will now have a  potential for
at least  the T-moduli (i.e. the size and shape 
of the squashed sphere) and of course a bulk 
cosmological constant. What we have is 
five dimensional gauged supergravity but the four dimensional world,
instead of being a further compactification of this, is viewed as 
a brane sitting in this five dimensional bulk. 
As discussed in 
\cite{sda}, there are various possibilities for 
getting flat four dimensional space on the brane 
without fine tuning\footnote{For a discussion 
of a  case which apparently does not involve a 
choice of integration constant but involves a 
bulk singularity see the first paper of \cite{ADS}. Apart from the 
difficulty of interpreting this naked singularity (which is common to
all the models in both papers of \cite{ADS} 
this model has the problem that the brane tension chosen is  
not renormalization group (RG) invariant. 
For comments on these models along these lines see
\cite{sda}, \cite{critique}.} none of which explain why the 
cosmological constant is zero. But unlike 
in the standard compactifications, where it is difficult 
to understand even how is the CC tuned to zero,  
in our case we believe there is a viable scenario 
whereby with a choice of integration constants and a compactification 
parameter (which is not determined by the ten dimensional theory) flat 
space four dimensional solutions can be obtained. 

Although our discussion is confined to an analysis of a 
five dimensional effective action coming from compactifying a ten
dimensional action for type IIB supergravity on a (squashed) 
sphere, we believe that 
is quite likely that there is a string construction 
in such a Ramond-Ramond (RR) background. In fact, recently there 
has been considerable progress in constructing such 
theories \cite{Berkovits}. Thus we think that the model 
that we propose here could be promoted to the complete string theory. 

Another issue that we should discuss here is that of singular 
brane configurations versus smooth configurations. 
We take the point of view that our branes are 
D-branes and/or orientifold planes so that as 
far as the low energy effective action is 
concerned their thickness is infinitesimal.\footnote{We stress that
although flat space considerations imply that gravity 
cannot be confined to a D-brane (as one wants for a 
one brane RS type scenario) in a warped background 
this is not necessarily the case.} 
Their transverse size in other words is set by the 
string scale and to the extent that we are 
working in the low energy
effective action we can ignore their thickness. 
Thus we will be looking for solutions to the 
bulk equations in the presence of these branes, 
which will act as singular source terms as  for instance in
section 3 of \cite{deW}. We believe that the considerations of
\cite{KL} where the authors look for brane scenarios that try to
mimic the Randall - Sundrum (RS) scenarios \cite{RS1} 
and/or \cite{RS2} without putting in sources 
(and finding a negative result) are not relevant 
to our construction. On either side
of our branes we will have smooth solutions to 
the supergravity equations which will then 
be matched at the position of the branes. It should also 
be pointed out here that a model for the real world 
would in fact require a brane which carried gauged 
degrees of freedom and so we really would
want to think in terms  of D-branes rather than smooth brane like
configurations. The picture that we have 
(i.e. basically that in \cite{RS1},\cite{RS2}) 
actually implies the physical existence of 
a brane (on which the standard model should live) 
placed in the bulk supergravity space. 

Our model has the explicit D-brane/orientifold planes
at the fixed points of the $S^1\over Z_2$ orbifold  
in the ($x^4$) fifth direction. We assume that  the ten 
dimensional dilaton is frozen by string scale dynamics. 
When the squashing 
modes are constant but at a certain critical point with  
non-zero values, the bulk is a 
$\cN =2$ background. Turning on the breathing mode 
$\vphi$ of the sphere
should not affect the supersymmetry. The supersymmetry of the D/orientifold 
3-plane would  be $\cN =1$. This should give  a solution of the 
RS type without fine tuning and with $\cN =1$ supersymmetry. 
After supersymmetry breaking however one does not expect 
a solution for arbitrary values of the radius of the $S^1$.  
In this case, in addition to the radius of the circle,
 there are 
also two integration constants from the metric factor 
and breathing mode and the constant coming from the compactification,
 - four constants in all 
that will be determined by the two pairs of matching conditions at each brane.
Thus all the constants are completely fixed. 

As stressed in \cite{sda}, this is of course 
not a solution to {\it the} cosmological constant 
problem since there is no explanation of why the 
integration constants are chosen to give the values 
that they need to have in order to get flat branes. 
Nevertheless we believe it is progress in that at 
least there is an explanation as to how one might 
get a zero CC in a string theoretic scenario 
after SUSY breaking.\footnote{While this paper 
was being prepared for publication a paper 
which discusses an alternate scenario  for two flat branes without fine tuning 
appeared. Our mechanism is replaced there with an 
assumption about supersymmetry in the bulk and 
on the ``Planck brane" \cite{Verlinde}. However we do not understand
how a fine tuning in the bulk potential is avoided in this case. }

It should also be pointed out that the brane world 
scenario appears to be  the only context, within 
string theory, that the idea of using the free parameter 
(from the point of view of the ten dimensional theory) in the
potential of gauged supergravity, to adjust 
the cosmological constant to zero, can be made use of. 
This is because this 
mechanism can only be used in type IIA or IIB theories 
(and orientifolds thereof) and 
the standard model/real world in such theories  
necessarily lives on a D-brane. 

In the next section we discuss a possible string 
theory set up in which the scenario we 
have in mind may be realized. In section 3
we discuss special bulk solutions to the five dimensional equations and 
in section 3.1 we display approximate general solutions with 
all the allowed integration constants that are necessary to obtain a 
solution with two branes. In section 4 we summarize the physics of
the construction and its significance, and then speculate on possible 
phenomenological applications.

%%%%%%%%%%%%%%%%%%%%%%%%%%%%%%%%%%%%%%%%%%%%%%%%%%%%%%%%%%%%
\section{The String Theory Setup}
%%%%%%%%%%%%%%%%%%%%%%%%%%%%%%%%%%%%%%%%%%%%%%%%%%%%%%%%%%%%

We wish to consider type IIB string theory 
compactified on a five sphere or squashed sphere. 
Typically this background is obtained by turning 
on the five form RR flux. We assume that 
string theory on such a background exists
perhaps along the lines discussed by Berkovits et al \cite{Berkovits}.
If this is indeed justified, it should follow that 
the entire D-brane machinery can be taken over.

Thus we look at $S^1/Z_2$ orientifold of the 
five dimensional theory resulting from the 
(squashed) sphere compactification with $S^1$ 
being a circle of radius $R$ and the $Z_2$ action 
being $x^4 \rightarrow -x^4$. The theory  is a 
five dimensional gauged supergravity which contains 
in addition to the dilaton-axion, 40 other 
scalars  some of which can be interpreted as squashing 
modes of the five sphere. These fields are 
massless in the limit when the gauge coupling is zero. 
In addition there is the so called (compactification scale) 
``breathing mode'' and of course the gauge fields which  
along with the two 2-form fields associated 
with the F and D strings will be set to zero 
in the following. Also we assume that the ten 
dimensional dilaton is frozen by string scale dynamics.

Let us first consider the round sphere compactification. 
The ten dimensional  metric is written as
$$ds_{10}^2=e^{2\a\vphi}ds_5^2+e^{2\b\vphi}ds^2(S^5)$$
with $\a ={1\over 4}\sqrt{5\over 3},~\b =-{3\over 5}\a$.
The ans${\ddot {\rm a}}$tz for the self-dual 5 - form  is 
\be\label{Hsolution}  H_{(5)}= 4{m }e^{8\a\vphi}\epsilon _{(5)}+4{m }\epsilon
_{(5)}(S^5), \ee 
where $\epsilon _{(5)}$ and $\epsilon _{(5)}(S^5)$ are the volume forms
of the non-compact and compact spaces respectively.
The effective five dimensional action is then \cite{Bremer}
\be\label{roundaction} S = \int d^5x\sqrt{-G_5}\bigl[{\cal R}-\shalf 
(\pa\vphi )^2+e^{16\a\vphi\over 5}R_5-8m^2e^{8\a\vphi}\bigr],\ee
where $R_5$ is the scalar curvature of the five sphere 
and we have set to zero all fields that are irrelevant to 
our discussion. 
Note \cite{Bremer} that
this action allows the $AdS_5$ solution with constant breathing mode
$\vphi =\vphi_0$ given by
\be\label{critphi} e^{{24\a\over 5}\vphi_0}={R_5\over 20m^2}\ee
and 
\be\label{} R_{MN}=-4m^2e^{8\a\vphi_0}{G_{5}}_{MN}.\ee

Now the fifth dimension is a $S^1\over Z_2$ orbifold. The 
fixed points would be orientifold planes and we place 
D-branes at one or other (or both) fixed points so 
as to have  two (composite) branes
at the ends of the fifth dimension. The picture is just like that in
the ten dimensional type IA situation analyzed by 
Polchinski and Witten \cite{pw}. In the original ten 
dimensional framework, the self dual five form field 
strength would satisfy in the presence 
of a collection of $n_0$ branes located near $x^4=0$ and
extending in the $x^1,x^2,x^3$ directions 
(counting -16 for the 
orientifold fixed plane), the equation $dH_{(5)} =n_0\tau\d (x^4)\w 
dx^4\w\e_5(S_5)$, where $\tau$ is the charge of a 
single brane. We take $x^0$ to be the time like direction and
$x^5,\cdots ,x^9$ the directions along the five sphere.
Substituting expression (\ref{Hsolution}) 
into this, gives $\pa_{x^4}m(x^4)=n_0\tau\d (x^4)$. The solution for 
$H_{(5)}$ now takes the form as before, i.e. 
(\ref{Hsolution}), but with $m$ 
being now a piece wise continuous function of $x^4$ with a jump
$\D m =n_0\tau$ at $x^4=0$ and a similar jump at $x^4=\pi R$.  
Using also the symmetry under $x^4\rightarrow -x^4$, we 
find the sort of situations illustrated in figs. \ref{f_warp4},
\ref{f_warp5}, \ref{f_warp6}.

%%%%%%%%%%%%%%%%%%%%%%%%%%%%%%%%%%%%%%%%%%%%
\FIGURE{\epsfig{file=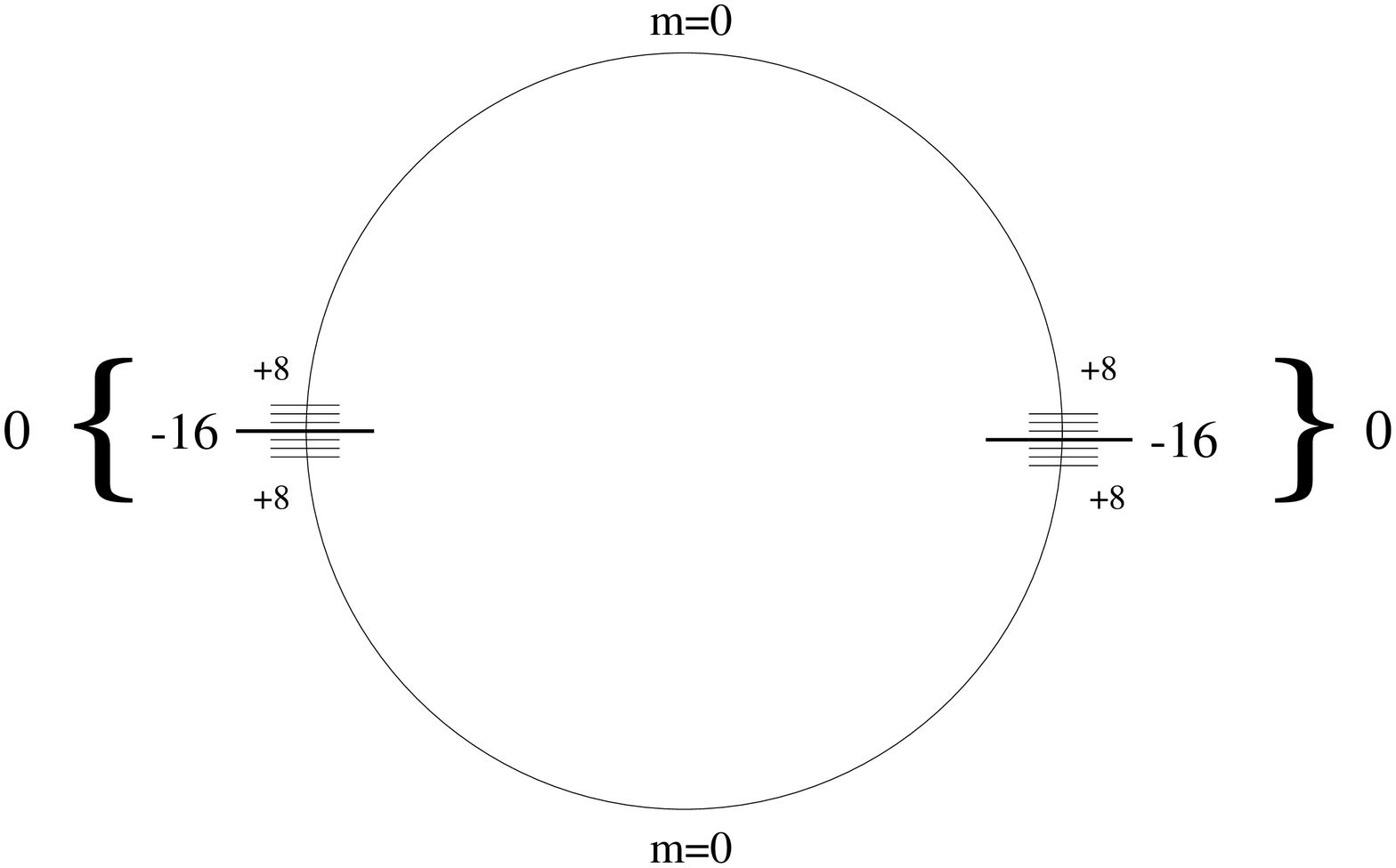,width=9.5cm}
        \caption{$IIB \;{\rm on} \;S^5$ brane world 1: $SO(16)\times SO(16)$.}
        \label{f_warp4}}
%%%%%%%%%%%%%%%%%%%%%%%%%%%%%%%%%%%%%%%%%%%%%

%%%%%%%%%%%%%%%%%%%%%%%%%%%%%%%%%%%%%%%%%%%%
\FIGURE{\epsfig{file=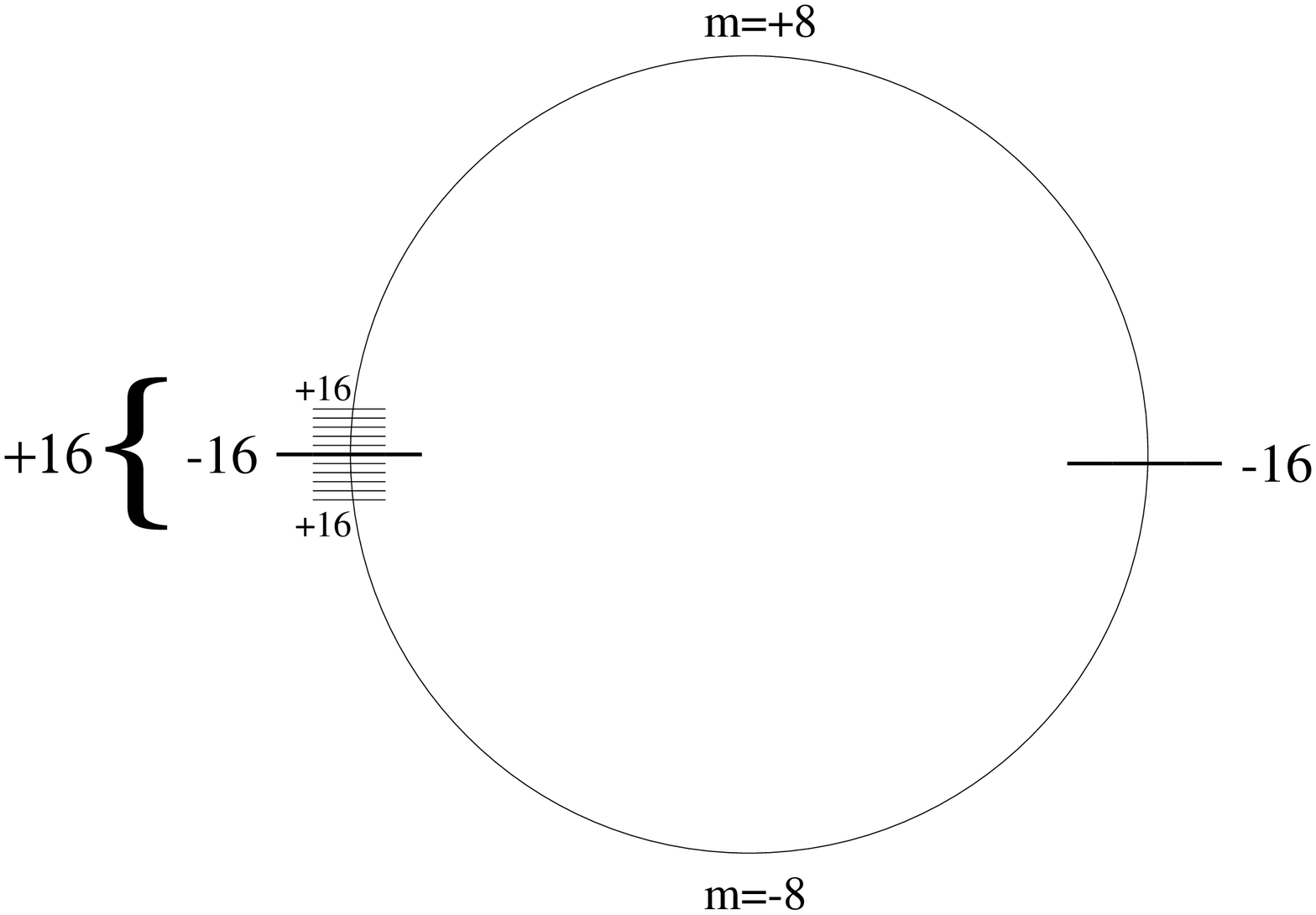,width=9.5cm}
        \caption{$IIB \;{\rm on} \; S^5$ brane world 2: $SO(32)$.}
        \label{f_warp5}}
%%%%%%%%%%%%%%%%%%%%%%%%%%%%%%%%%%%%%%%%%%%%%

%%%%%%%%%%%%%%%%%%%%%%%%%%%%%%%%%%%%%%%%%%%%
\FIGURE{\epsfig{file=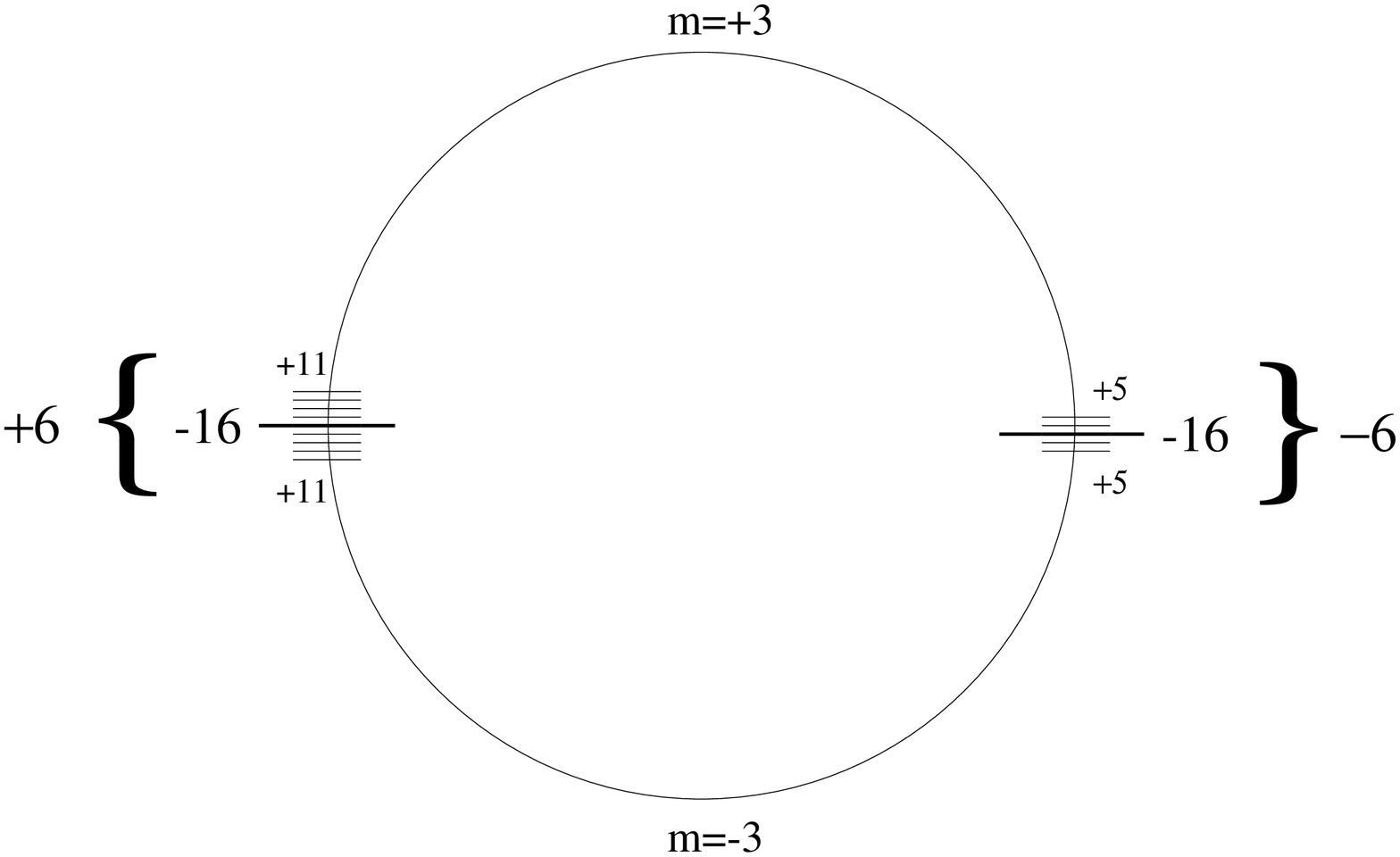,width=9.5cm}
        \caption{$IIB \;{\rm on} \; S^5$ brane world 3: $SO(22)\times SO(10)$.}
        \label{f_warp6}}
%%%%%%%%%%%%%%%%%%%%%%%%%%%%%%%%%%%%%%%%%%%%%

Note that the quantization of $H_5$ also follows directly from its
coupling 
to the $D3$ brane. Consistent coupling requires by a standard argument
$\tau\int_{S_5} H_5=2\pi n$ where $n$ is an integer and $\tau =2\pi
M_s^4$ is the $D_3$ brane charge in terms of the string scale $M_s$. Using
equation (1) for $H_5$ and $\int_{S_5}=\pi^3r^5$ we then get
$$ m =n\left (1\over M_s^44\pi ^3r^5\right ).$$
The units for $m$ given in the figures is then the quantity in parantheses
in the above equation.

In the above, we should point out that we have assumed 
not only that there is a valid string description as 
in Berkovits et al \cite{Berkovits} but also that the 
corresponding five dimensional
orientifolds have charges $-16$. These values are just 
given for illustrative purposes and of course if the general scenario is
valid with however different charges, then the examples 
would have to be modified accordingly. We should also 
point out that alternatively, one might try to justify  such a picture 
by considering the following procedure.
First, compactify type I string theory on a five torus. 
Taking the T-dual picture one has a type IIB orientifold 
with $2^5$ orientifold 3-planes at the fixed points of 
the $Z_2$ orbifold symmetry of the dual torus and 32 
D3-branes to cancel the tadpoles.
Now let the size of the dual torus go to infinity 
while keeping some number of the D-branes at the fixed point at the
origin. Now adding the point at infinity should give presumably an
$S^5/Z_2$. It is not 
quite clear to us which compactification makes sense 
in the complete string theory but given that 
the 5D gauged supergravity models appear to come from 
five sphere or squashed sphere compactifications we will
just consider these, assuming that a theory along the lines of Berkovits et al 
\cite{Berkovits} will allow our scenario.

As we will see below, 
the bulk is taken to be at the ${\cal N}=2$ critical point 
of the gauged supergravity potential. In the presence of 
the branes this is then broken down to ${\cal N}=1$ so 
that one may have a phenomenologically acceptable 
model on one or the other brane.
Let us pick the one at $x^4=0$ to situate the standard model. 
The other one can then play the role of a hidden sector 
(as in the Horava-Witten theory \cite{HW}) where supersymmetry 
can be broken and then communicated to the visible brane. 
The only new point here is that
any cosmological constant that is generated as a result 
of the SUSY breaking is compensated  by adjustment of 
integration constants and $R_5$. 

It should be remarked here that $R_5$ is  
on the same footing as the other integration constants 
(which occur in the five dimensional theory) from the 
point of view of the ten dimensional theory. It is not a modulus. 
The relevant modulus is the field $\vphi$. $R_5$ is, from 
the point of view of the ten dimensional theory, and  
presumably of string theory, an integration constant 
appearing in the ansatz for the ten dimensional metric. 
For compactification on a Ricci flat metric such as 
a Calabi-Yau metric, this constant can be absorbed into $\vphi$. 
In our case as long as $m$ is non-zero i.e. the 
potential is not runaway, $R_5$ cannot be absorbed into the modulus $\vphi$.
We stress again that this is not a mechanism that was available with
standard compactification scenarios 
(Calabi-Yau, Orbifold etc).\footnote{Recently,
other warped compactifications, in the context of M/F theory, have
appeared \cite{M} where a RS type scenario with 
explicit branes is constructed. However for these  
Calabi-Yau compactifications with G-flux, it is not possible to use 
our mechanism, since the compact space is Ricci flat, 
so one does not have the required freedom 
in the choice of constants to cancel the CC.}

The next issue is what is the appropriate 5D effective theory. Our
choice is the ${\cal N}=2$ supersymmetric vacuum of \cite{vacuum} 
which corresponds to two non trivial scalars relaxed to their critical
values plus a non trivial breathing mode, 
a generalization of the model with one squashing
mode and a non trivial breathing mode model of \cite{Bremer}. The scalar
potential of this model is expected to have a contribution from the
curvature of the compact space and a number of contributions from 
non-trivial 5-form and 3-form fluxes. For simplicity, we will truncate the
potential only to the curvature term and the 5-form flux term since
these are sufficient to demonstrate our point. A more complete treatment
should include all the other terms. The potential is then just as in the
round sphere case (\ref{roundaction}), and it is:
\be V(\vphi)=-R_5e^{a\vphi}+8{m }^2e^{b\vphi}.\label{eq:Vbr}\ee 

%%%%%%%%%%%%%%%%%%%%%%%%%%%%%%%%%%%%%%%%%%%%%%%%%%%%%%%%%%%%%%%%%%%%%%%%%%%
\section{Flat Domain Wall Solutions - Brane Worlds}
%%%%%%%%%%%%%%%%%%%%%%%%%%%%%%%%%%%%%%%%%%%%%%%%%%%%%%%%%%%%%%%%%%%%%%%%%%%

The equations of motion corresponding to the action (\ref{roundaction})
with $\vphi=\vphi (x^4)$ 
and the flat four dimensional domain wall $5D$ metric  
\be
ds_5^2=e^{2A(x^4)}\eta_{\mu\nu}dx^{\mu}dx^{\nu}+{(dx^4)}^2,\label{eq:metric}\ee
where $\eta_{\mu \nu}=diag(-1,+1,+1,+1)$,
are
\be \vphi ''+4A'\vphi '={{{\partial 
V(\vphi)}\over {\partial {\vphi}}}},\label{eq:Beq1}\ee
\be A''=-{1\over 6}\vphi'^2,\label{eq:Beq2}\ee
\be A'^2=-{1\over 12}V(\vphi)+{1\over 24}\vphi'^2.\label{eq:Beq3}\ee
The prime denotes differentiation with respect to $x^4$.
We emphasize that our domain wall ansatz is that 
we are looking for solutions with flat four dimensional slices of the
five dimensional geometry.

It has been shown that by introducing a ``superpotential'' 
$W$, the second order differential equations reduce to \cite{deW}, \cite{ST}: 
\be \vphi'={{{{\partial W(\vphi)}\over 
{\partial {\vphi}}}}},\label{eq:eq1}\ee 
\be A'=-{1\over 6}W,\label{eq:eq2}\ee
\be V(\vphi)={1\over 2}\Bigl({{{\partial W}\over {\partial
{\vphi}}}}\Bigr)^2-{1\over 3}W^2.\label{eq:VW}\ee
This is a system of decoupled first order differential equations instead of the
coupled second order differential equations that we had before.
A particular ansatz for $W$ that satisfies the last condition 
for $V(\vphi)$ of the form (\ref{eq:Vbr}), is
\be W=p_1e^{{a\over 2}\vphi}+
p_2e^{{b\over 2}\vphi}\label{eq:W},\ee
with 
\be p_1=\pm 2\sqrt{{{2R_5}\over {a(b-a)}}}\;\;\;\; {\rm and}\;\;\;\; 
p_2=\pm 2\sqrt{{{16m^2}\over {b(b-a)}}}.\ee
This ansatz essentially solves the system, 
but it seems that we have to pay a price.
Namely, we have lost one integration constant 
because (\ref{eq:W}) is a particular solution of (\ref{eq:VW}) 
and does not have the integration constant 
that a general solution should have. 

Note that the original second order system apparently 
has four integration constants. However,
(\ref{eq:Beq3}) is a constraint equation which 
gives one relation between these constants. 
Also the zero mode of $A$ can always be 
absorbed in a rescaling of coordinates. 
Thus the original system has two independent 
integration constants. As stressed by \cite{deW}, 
the first order system must be
completely equivalent to the second order system 
and the parameter space should have the same dimension. 
This is of course possible only if
we use the general solution for $W$ in which case 
the first order system will also have two independent 
integration constants, one in the solution for $W$ 
and one in the solution for $\vphi$. Of course if
we work with the particular solution (\ref{eq:W}), 
we would be restricting ourselves to a one parameter 
subspace of the solution space.
We will come back to the problem of the lost integration constant later. 

For the moment, we have to solve the following system of first order
differential equations:
\be \vphi'={1\over 2}ap_1e^{{a\over 2}\vphi}+
{1\over 2}bp_2e^{{b\over 2}\vphi},\label{eq:diff1}\ee
\be A'=-{1\over 6}p_1e^{{a\over 2}\vphi}-
{1\over 6}p_2e^{{b\over 2}\vphi}\label{eq:diff2}.\ee
Integrating (\ref{eq:diff1})  gives the
Lerch transcendent \cite{clp} as the solution,
\be r+c_1^{\vphi}={2\over {ap_1}}
\sum_{n=0}^{\infty}{{(-{{bp_2}\over {ap_1}})^n}\over {(a-b)n+a}}
\Bigl[1-e^{\bigl(n(b-a)-a\bigr)\vphi}\Bigr],\label{eq:Lerch}\ee
which is quite hard to invert for general $\vphi$,
so as to obtain an exact solution for the warp factor. 
It is much easier to work in certain interesting
limits:
\begin{itemize}
\item $\vphi\rightarrow \vphi_0$.

Let us first solve (\ref{eq:diff1}) and (\ref{eq:diff2}) for
$\vphi=\vphi_0$, the value of $\vphi$ at the 
critical point of $V$. We have $\vphi'=0$ and therefore ${\partial W\over
{\partial \vphi}}=0$, which implies through $(\ref{eq:VW})$ that this
vacuum characterizes a critical point 
(in fact a minimum) of the scalar potential. 
Furthermore,
from (\ref{eq:diff2}) we see that it corresponds to an exact $AdS$
vacuum, since $A$ has a linear dependence in $x^4$. The condition,
therefore, that determines $\vphi_0$, is
\be ap_1e^{{a\over 2}\vphi_0}+bp_2e^{{b\over 2}\vphi_0}=0.\label{eq:AdS}\ee
Let us now solve the equations for a vacuum that is near the
$\vphi=\vphi_0$ vacuum. For that, we assume
\be \vphi=\vphi_0+\epsilon f(x^4),\ee
with $\epsilon $ a small number. Substituting this ansatz into
(\ref{eq:diff1}), we find that $f$ satisfies 
the differential equation $f'=4kf$ and therefore
\be \vphi=\vphi_0+c_1^{\vphi}e^{4kx^4}, \;\;\;\;\;\;
k^2={1\over 32}{a(b-a)}R_5e^{a\vphi_0},\label{eq:phi}\ee
where $k$ and $p_1$ are of opposite signs and 
$c_1^{\vphi}$ is the integration constant with $\epsilon$ absorbed in it.
Solving (\ref{eq:diff2}) for $A$, yields the (almost $AdS$)  warp factor:
\be A(x^4)=kx^4+{\cal O}(\epsilon^2).\label{eq:AA}\ee
 
The first term in the above, corresponds to the $AdS$ part. The 
solution (\ref{eq:phi}) and (\ref{eq:AA}), however, is valid only for large
$|x^4|$. In particular, if $k<0$ 
(that is if $p_1>0$) then the solution is valid when
$x^4\rightarrow +\infty$ and if $k>0$ ($p_1<0$) then the solution is valid when
$x^4\rightarrow -\infty$. Finally, (\ref{eq:AdS}) implies that $p_1$ and
$p_2$ must come with opposite signs.
\item $\vphi\rightarrow +\infty$.

In this limit, we have     
\be \vphi '\simeq {1\over 2}bp_2 e^{{1\over 2}b\vphi },\ee
which can be integrated to give 
\be \vphi = -{2\over b}\Bigl[\ln{\bigl(-{b^2\over
4}p_2x^4}+c_1^{\vphi}\bigr)\Bigr],\label{eq:rinfphi}\ee
where $c_1^{\vphi}$ is an integration constant.
The above solution is valid for 
 $-{b^2\over 4}p_2x^4+c_1^{\vphi}\rightarrow 0^+$. 
The solution for the warp factor turns out to be 
\be e^{2A(x^4)}=\Bigl({-{b^2\over 4}p_2x^4+
c_1^{\vphi}}\Bigr)^{{4}\over {3b^2}},\ee
up to an (irrelevant) overall integration constant.
Thus in this limit, we have $e^{2A}\rightarrow 0^{+}$.  
\item $\vphi\rightarrow -\infty$.

In this limit, we have     
\be \vphi '\simeq {1\over 2}ap_1 e^{{1\over 2}a\vphi },\ee
giving 
\be \vphi = -{2\over a}\Bigl[\ln{\bigl(-{a^2\over
4}p_1x^4}+c_1^{\vphi}\bigr)\Bigr],\label{eq:rin2phi}\ee
which implies that 
 $-{a^2\over 4}p_1x^4+c_1^{\vphi}\rightarrow +\infty$
for any finite $c_1^{\vphi}$.

The warp factor is 
\be e^{2A(x^4)}=\Bigl({-{a^2\over 4}p_1x^4+
c_1^{\vphi}}\Bigr)^{{4}\over {3a^2}},\ee
up to the usual overall constant.
Thus, we have $e^{2A}\rightarrow +\infty $ in this limit. 
\end{itemize}

As was already noticed in \cite{clp}, the above solution has two separate
branches, one which corresponds to $\vphi \in (\vphi_0, +\infty)$ for 
$x^4\in (-\infty, \; {4\over {b^2p_2}}c_1^{\vphi})$ and one which
corresponds to $\vphi \in (\vphi_0, -\infty)$ for 
$x^4\in (-\infty, +\infty)$. We will see later that the separation of
the solution in two distinct branches is probably
special to our specific choice (\ref{eq:W}) of $W$
and that this choice is equivalent to choosing a gauge where $c_2^{\vphi}=0$. 
The first branch is singular at $x^4={4\over {b^2p_2}}c_1^{\vphi}$
because the warp factor vanishes, so we neglect it. 

%%%%%%%%%%%%%%%%%%%%%%%%%%%%%%%%%%%%%%%%%%%%
\FIGURE{\epsfig{file=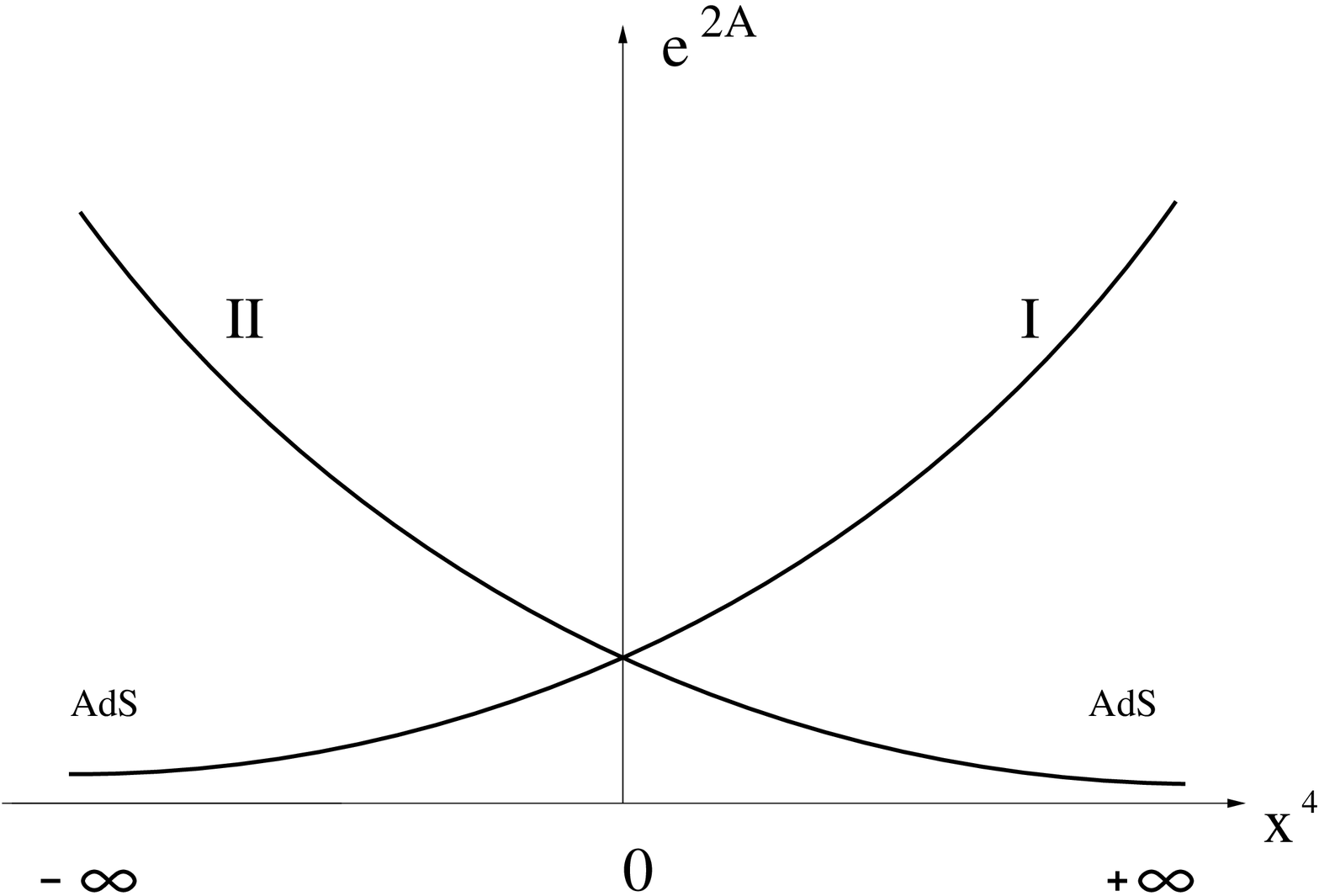,width=9.5cm}
        \caption{Solutions for branes (\cite{clp}): 
Warp factor as a function of $x^4$; for 
solution $I$, the warp factor behaves as 
an exponential ($AdS$) at $-\infty$ and as
a polynomial around $+\infty$. Solution $II$ can be obtained from
solution $I$ by $x^4\rightarrow -x^4$.}
        \label{f_warp2}}
%%%%%%%%%%%%%%%%%%%%%%%%%%%%%%%%%%%%%%%%%%%%%

As an example, 
in fig. \ref{f_warp2}, we plot the first branch solutions ($I$ and $II$) 
for the warp factor found in \cite{clp}. They correspond to $p_1<0$ and
$p_1>0$ respectively, i.e. solution $II$ can be obtained from solution
$I$ by a reflection around the origin $x^4=0$.
Solution $I$ corresponds to patching together in a smooth way the above found
solutions for $\vphi\rightarrow \vphi_0$ and $\vphi\rightarrow -\infty$
in which case the singularity encountered in the 
$\vphi\rightarrow +\infty$ regime is avoided.
Solutions $I$ and $II$ individually do not possess reflection
symmetry around the origin, but they are the ones appropriate for
constructing models with explicit branes inserted in the bulk. 
More specifically, we can construct a brane world scenario as follows: 
Consider the solutions of fig. \ref{f_warp2} and choose a region 
in the $x^4$ coordinate, symmetric around $x^4$.
The region is, say, the interval $[-\pi R,+\pi R]$. 
If $-\pi R <x^4 < 0$, take solution $I$ as it is
and if $0\le x^4 \le +\pi R$, take solution $II$ to get 
a $Z_2$ symmetric situation. The orbifold will then be the region 
$0<x^4<\pi R$. A one brane world can be
obtained by taking $R$ to infinity.  
This construction, for the warp factor, is illustrated in fig.
\ref{f_warp3}. 

%%%%%%%%%%%%%%%%%%%%%%%%%%%%%%%%%%%%%%%%%%%%
\FIGURE{\epsfig{file=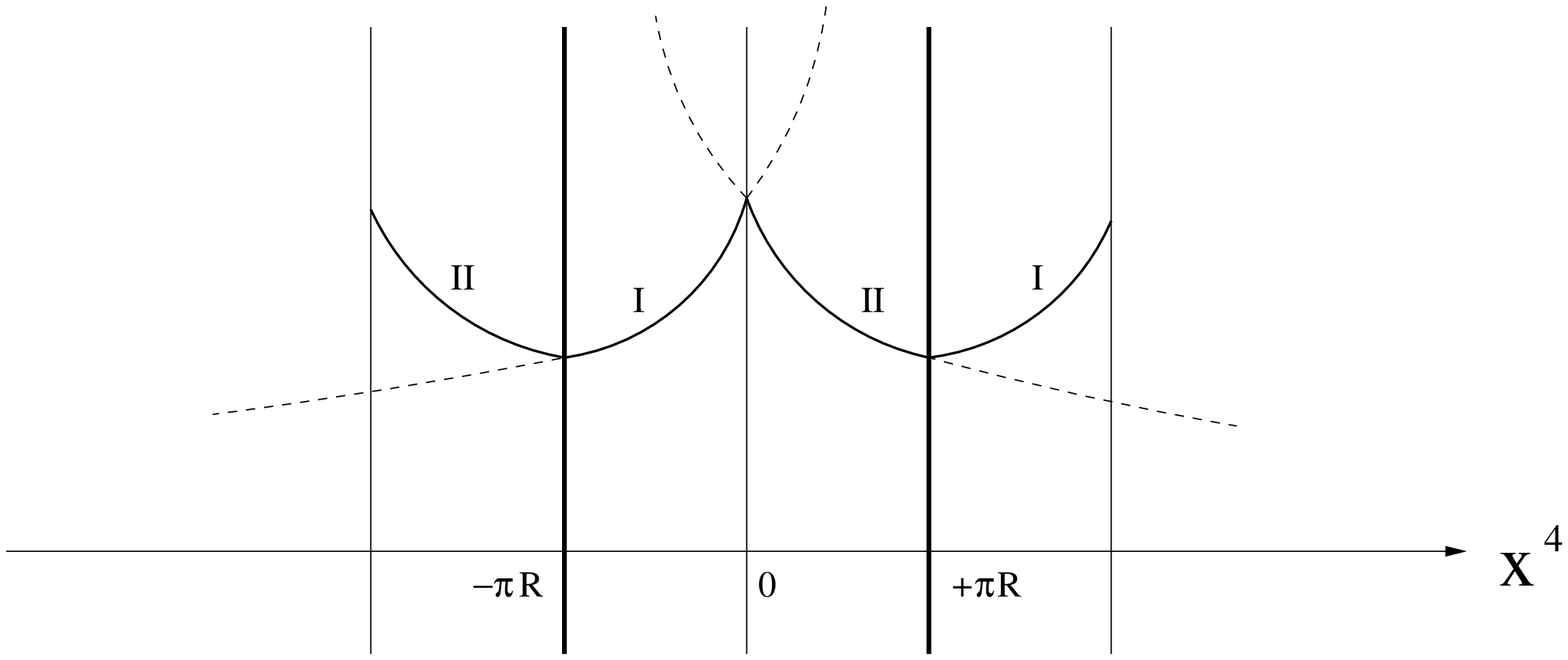,width=16cm}
        \caption{The orbifold construction.}
        \label{f_warp3}}
%%%%%%%%%%%%%%%%%%%%%%%%%%%%%%%%%%%%%%%%%%%%%

With the $D$3-branes present, the $5$ dimensional action is modified to:
\begin{eqnarray} S(\vphi)=\int {d^5x\sqrt{-G_5}\Bigl[    
{\cal R}-{1\over 2}(\partial \vphi)^2-V(\vphi) \Bigr]} 
-\int {d^4x\sqrt{-G^{(1)}_{4}}T^{(1)}(\vphi)}- \nonumber \end{eqnarray}
\begin{eqnarray}-\int {d^4x\sqrt{-G^{(2)}_{4}}T^{(2)}(\vphi)},
\label{eq:actionbrain}\end{eqnarray}
where $G^{(i)}_{4\mu\nu}$ is the induced metric 
on the $(i)$'th brane by $G_{5\mu\nu}$, and
for our metric it is simply $G^{(i)}_{4\mu\nu}=\eta_ {\mu \nu}$, and 
$T^{(1)}(\vphi)$ and $T^{(2)}(\vphi)$ are the 
tensions of the branes. The dependence of $T$ on 
$\vphi$ is kept arbitrary since it will in general be 
affected by quantum effects on the brane when supersymmetry is broken.
The action for the branes is written in the static gauge
so that the embedding functions are 
$x^{\mu}(\xi)=\xi^{\mu}$, $\mu=0,\cdots ,3$ and we ignore their
fluctuations. Also of course we have set all other fields on the 
brane to the minimum of the quantum effective action so 
that what we have kept is just an effective description 
of the ground state of the theory on the brane.

Now, in addition to the equations of motion, we have 
to satisfy the jump conditions
at $x^4=0$ (and $x^4=\pi R$ in the two brane case), 
which constitute the connection between
the two solutions at opposite sides of the brane(s). They are:
\be 2A'(x^4)=+{1\over 6}T^{(1)}(\vphi (x^4))\mid_{x^4=0},\label{eq:jump1}\ee 
\be 2A'(x^4)=-{1\over 6}T^{(2)}(\vphi (x^4))
\mid_{x^4=\pi R},\label{eq:jump2}\ee    
\be 2\vphi '(x^4)=-{\partial T^{(1)}\over {\partial {\vphi}}}
(\vphi (x^4))\mid_{x^4=0},\label{eq:jump3}\ee 
\be 2\vphi '(x^4)=+{\partial T^{(2)}\over {\partial {\vphi}}}(\vphi
(x^4))\mid_{x^4=\pi R}.\label{eq:jump4}\ee 

%%%%%%%%%%%%%%%%%%%%%%%%%%%%%%%%%%%%%%%%%%%%%
\subsection{More General Solutions}
%%%%%%%%%%%%%%%%%%%%%%%%%%%%%%%%%%%%%%%%%%%%%

In a two-brane world, we need both nontrivial
integration constants of the second order equations of motion
(\ref{eq:Beq1}), (\ref{eq:Beq2}) and (\ref{eq:Beq3}),
because we have only $R_5$ and the size of the orbifold $R$ to satisfy
four jump conditions. 
One might think a solution with two branes and a
pure $AdS$ bulk is possible \cite{RS1}, 
since there are only two non trivial jump
conditions to satisfy (one for each brane) and two available constants,
$R_5$ and $R$. However, as was pointed out in \cite{sda}, the two branes
have equal and opposite tensions at every point of the RG flow and the
tension for a flat brane solution is fixed only in terms of $R_5$. The
orbifold size does not enter the jump conditions and therefore it can
not be used. In order to have a solution without fine tuning, we need a
nontrivial scalar field in the bulk. But then, our initial choice 
of $W$ does not provide us with the two required integration constants,
since as we explained, we have lost one integration constant 
in the choice (\ref{eq:W}). 
We therefore look for more general solutions to the
equations of motion.

We make the ansatz 
\be \vphi=\vphi_0+\epsilon f_1(x^4)+\epsilon^2 f_2(x^4)+\epsilon^3
f_3(x^4)+\cdots \ee
\be A=kx^4+\epsilon g_1(x^4)+\epsilon^2 g_2(x^4)+\epsilon^3 g_3(x^4)+\cdots \ee
with $k>0$, and we substitute it into 
(\ref{eq:Beq1}), (\ref{eq:Beq2}) and (\ref{eq:Beq3}). We obtain an
infinite set of differential equations. We show the result 
for (\ref{eq:Beq1}) and (\ref{eq:Beq2}), up to order $\epsilon ^3$:
\be f_1''+4kf_1'+pf_1=0,
\;\;\;\; g_1''=0 \label{eq:expan1}\ee
\be f_2''+4kf_2'+pf_2-{1\over 2}a(a^2-b^2)R_5e^{{1\over
2}a\vphi_0}f_1^2+4f_1'g_1'=0,
\;\;\;\; g_2''+{1\over 6}f_1'^2=0 \label{eq:expan2}\ee 
\begin{eqnarray} f_3''+4kf_3'+pf_3-a(a^2-b^2)R_5e^{{1\over 2}a\vphi_0}f_1f_2-
{1\over 6}a(a^3-b^3)R_5e^{{1\over 2}a\vphi_0}
f_1^3+ \nonumber \end{eqnarray}
\begin{eqnarray}+4f_1'g_2'+4f_2'g_1'=0,\;\;\;\; g_3''+{1\over
3}f_1'f_2'=0,\label{eq:expan3}\end{eqnarray}
where $p=-32k^2$.
The above pattern (that presumably continues),
suggests that we can write the general solution for the field
$\vphi$ as:
\be \vphi (x^4)=\vphi_0 + c_1^{\vphi}e^{4kx^4}
+c_2^{\vphi}e^{-8kx^4}+P(x^4,c_1^{\vphi},c_2^{\vphi}),\label{eq:sol}\ee
where the part of the expression with the exponentials is the 
general solution to the equation 
$f''+4kf'+pf=0$ and $P(x^4,c_1^{\vphi},c_2^{\vphi})$ is 
some function of $x^4$ containing also the two integration constants
and $\epsilon$ has been absorbed into the integration constants.
By looking at (\ref{eq:expan1}), (\ref{eq:expan2}) and
(\ref{eq:expan3}), one can see that the solution is a small
deviation from $AdS$ in two cases. One case is when 
$x^4\rightarrow -\infty$. Then, the expansion in $\epsilon$
makes sense only if $c_2^{\vphi}=0$ and therefore this case is just the
solution (\ref{eq:phi}). The second case is when $x^4\simeq 0$ and 
$c_1^{\vphi},c_2^{\vphi} << 1$. This is  
a satisfactory solution with two
explicit integration constants as long as the jump conditions allow 
them to be small. Another observation we can make here is that when
$c_2^{\vphi}\ne 0$, it is not clear that $\vphi$ has two separate
branches. Nevertheless,
we can still apply the method for the construction of the orbifold
with the only caveat that if the jump conditions require
integration constants of order one or larger, we still do not have an
explicit solution.  

We will now find a solution to the second order 
equations of motion in the $\vphi<\vphi_0$ region which will be valid
for a larger range of the integration constants.
Define $H\equiv A'$ and denote ${\partial \over {\partial \vphi}}$ with
a dot. Also, for simplicity we will assume that $R_5=8m^2=1$, even
though we have to keep in mind that $R_5$ is really a parameter determined by
the integration constants. 
Combining equations (\ref{eq:Beq2}) and
(\ref{eq:Beq3}), we obtain the equation ${\dot H}^2={1\over 36}({24H^2+2V})$. 
Let us assume that $24H^2>>2\vert V \vert$. 
We will verify soon that this is a valid
assumption in the regime of interest. Then, we can easily solve for $H$:
$H=e^{\sqrt{2\over 3}\vphi}$, where we have chosen the positive branch
of the square root. Using the value of $\vphi_0$ from the minimization
of $V$, we deduce that $2V$ could be consistently dropped provided that 
\be {1\over 12} e^{a\vphi}\Bigl(1-{a\over b}
e^{(b-a)(\vphi -\vphi_0)}\Bigr)<< 
e^{2{\sqrt{2\over 3}\vphi}}.\label{eq:cond1}\ee
Since our solution corresponds to $\vphi \in (-\infty,
\vphi_0)$, a numerical estimate tells us that 
the above is true if, approximately, $-3 < \vphi < -0.6$
(for $R_5=8m^2=1$, we get from (\ref{eq:AdS}) $\vphi_0\simeq -0.6$).
Next, using the above solution for $H$, equation (\ref{eq:Beq1}) becomes 
$\vphi''+4e^{\sqrt{2\over 3}\vphi}\vphi'-{\dot V}=0$. In the
$\vphi < \vphi_0$ region the potential $V$ 
approaches zero exponentially and its
shape in this regime is
rather flat, which means that ${\dot V}\simeq 0$. Indeed, one can
neglect ${\dot V}$ in the equation for $\vphi$ provided that
\be \vert -ae^{a\vphi}+be^{b\vphi}\vert << \; 
4e^{{\sqrt{2\over 3}\vphi}}\vert \vphi'\vert
\;\;\; {\rm and} \;\;\; \vert \vphi'' \vert .\label{eq:cond2}\ee
Then we can solve for $\vphi$ and we obtain 
\be \vphi (x^4)={\sqrt {6}\over 2}
\ln {{c_2^{\vphi}}\over {1+12e^{{{\sqrt {6}\over 3}}c_2^{\vphi}
(c_1^{\vphi}+x^4)}}}+c_2^{\vphi}(c_1^{\vphi}+x^4)+
{\sqrt {6}\over 4}\ln{6}.\label{eq:vphisol} \ee
By computing $\vphi '$ and $\vphi ''$ from the above expression, one can
verify that (\ref{eq:cond2}) is true in the relevant range of $\vphi$
for a large range of $c_1^{\vphi}$ and $c_2^{\vphi}$, 
when $x^4\sim {\cal O}(0-10)$.  
\footnote{(\ref{eq:cond2}) is in fact true for any finite $x^4$ for 
appropriate choice of the order of magnitude of the
integration constants, but the regime around
$x^4=0$ is the interesting one since it is where the branes are located.}

For $\vphi>>\vphi_0$, we can take similar steps. We can again assume for
simplicity that $R_5=8m^2=1$. The derivative of the
warp factor is obtained by solving the equation ${\dot H}={\sqrt{2}\over
6}e^{{b\over 2}\vphi}$, which yields $H={\sqrt{2}\over {3b}}e^{{b\over
2}\vphi}$ up to an irrelevant integration constant. The equation
for $\vphi$ becomes $\vphi''=be^{b\vphi}$, where we have ignored the smaller
terms. The solution for $\vphi$ is then
\be \vphi (x^4)={\sqrt{15}\over 10}\ln{\Bigl[-{c_1^{\vphi}\over 2}\bigl
[1+\tan{\Bigl({-c_1^{\vphi}}{{5}\over 3}
(x^4+c_2^{\vphi})^2\Bigr)}\bigr]\Bigr]}.\ee  
Finally, we assure that one can solve the equations of motion in an
analogous fashion for arbitrary $R_5$.

We can now construct the orbifold as in fig.\ref{f_warp3}
and satisfy the four
jump conditions. An interesting fact is that to do so, 
along with $c_1^{\vphi}$ and $c_2^{\vphi}$, 
we have  to use $R_5$ and $R$, which
provides us with a RG scale dependent determination of the 
$5D$ dilaton and the orbifold size.

%%%%%%%%%%%%%%%%%%%%%%%%%%%%%%%%%%%%%%%%%%%%%%%%%%%%%%%%%%%%%%%%%%%%%%%%%%%
\section{Conclusions}
%%%%%%%%%%%%%%%%%%%%%%%%%%%%%%%%%%%%%%%%%%%%%%%%%%%%%%%%%%%%%%%%%%%%%%%%%%%

In this paper we have suggested a string theoretic framework
for a two brane scenario. The main issue we have addressed is
the possibility of obtaining a flat four dimensional world
on the branes. The supersymmetric case is a degenerate one 
in which the matching conditions 
(equations (\ref{eq:jump1}-\ref{eq:jump4}))  are 
automatically satisfied. When supersymmetry is broken however
these conditions are non-trivial. The solutions to the equations
of motion and constraint have two independent integration constants.
In addition there is the distance between the branes yielding in
general three adjustable constants in all. As was observed in
\cite{deW} this would mean that there would have to be a tunable
parameter either in the bulk potential or the brane tension.

The main observation of this paper is that there is a possible
string construction in which the appearence of such a tunable
parameter in the five dimensional theory is manifest. Indeed
at the ten dimensional level there is no tuning, the parameter
appears in the compactification and is in this sense an integration
constant. Its appearence is related to the fact that we are dealing 
with a Ricci non-flat compactification 
(which gives a potential with a critical 
point for the breathing mode). 

Solutions in the presence of supersymmetry breaking necessarily
must have the complete set of free parameters that we have enumerated
above. In particular the solutions for the warp 
factor and the breathing mode must contain the 
two independent integration constants. Unfortunately 
it was not possible to obtain exact analytic
solutions displaying this but we have shown how 
to obtain approximate solutions that contain the two integration constants.
In other words we have demonstrated the existence of (approximate)
solutions that can be used to support flat branes after supersymmetry
breaking, justifying the picture illustrated in figure 5.

There is one major unsolved problem though.
What we have done strictly speaking is to demonstrate the existence
of a flat two brane scenario after supersymmetry 
breaking in the context of type IIB supergravity. 
In the paper we have suggested
that this ought to imply that there is a corresponding microscopic
i.e. string theoretic implementation but this was not 
explicitly demonstrated. In particular in the supersymmetric
situation, the D-brane tension in five dimensions should be 
obtained from the usual formula for D-brane 
tension in ten dimensions. However, as was demonstrated 
in recent papers \footnote{These appeared
after the first version of this paper was published 
on the e-print archive hep-th.} \cite{BKV}\cite{DLS},  
five dimensional supersymmetry requires that the tension $T(\vphi )$ is
essentially  the superpotential $W$ of equation (\ref{eq:W}). It is not
at all clear how this comes from the usual ten dimensional action 
for the D3 brane though it is  possible that this action needs to
be modified for string theory in the presence of 
RR flux and Ricci-non-flat compactifications. 
Perhaps one should expect a simple relation 
to the 10 dimensional tension only at the maximally supersymmetric
point, the $\cal N$=8 critical point (\ref{critphi}). 
At this point one finds that the tension
in five dimensions $T(\vphi_0)={3\over 4}T_{D_3}$, a relation that was
first found by \cite{Kraus}. However this apparently can be justified
from the string point of view \cite{DLS}. Perhaps this is all that is
needed, but the matter is not entirely clear 
to us and is currently under investigation.

The phenomenological importance of our scenario is that 
we can  have a ${\cal N} =2$ bulk five dimensional 
theory (by compactifying on a squashed five sphere as 
for example in \cite{squashed},
so that the brane theory would be ${\cal N} =1$ in four dimensions.
The implication in four dimensions is that
we have an adjustable constant (essentially an 
integration constant) in the superpotential 
that can be used to set the cosmological constant 
to zero after supersymmetry breaking 
(say by gaugino condensation). If this scenario 
can be completely justified (i.e. if the problem 
mentioned above can be solved) then it would be 
the first time that a string theoretic justification 
could be given for adding an adjustable constant to the superpotential.

\bigskip

%%%%%%%%%%%%%%%%%%%%%%%%%%%%%%%%%%%%%%%%%%%%%%%%%%%%%%%%%%%%%%%%%%%%%%%%%%%
\acknowledgments  
%%%%%%%%%%%%%%%%%%%%%%%%%%%%%%%%%%%%%%%%%%%%%%%%%%%%%%%%%%%%%%%%%%%%%%%%%%%

This work is partially supported by the Department of
Energy contract No. DE-FG02-91-ER-40672.

\bigskip

%%%%%%%%%%%%%%%%%%%%%%%%%%%%%%%%%%%%%%%%%

%%%%%%%%%%%%%%%%%%%%%%%%%%%%%%%%%%%%%%%%%%%%%%%%%%%%%%%%%%%%%%%%%%%%%%%%%%%%

%%%%%%%%%%%%%%%%%%%%%%%%%%%%%%%%%%%%%%%%%%%%%

\begin{thebibliography}{999}
%%%%%%%%%%%%%%%%%%%%%%%%%%%%%%%%%%%%%%%%%

\bibitem{gsw}
M.~B.~Green, J.~H.~Schwarz and E.~Witten,
``Superstring Theory''. Vol. 1 and 2 
{\it  Cambridge, Uk: Univ. Pr.} ( 1987) 469 P.

\bibitem{bda}R.~Brustein and S.~P.~de Alwis,
``String universality'',
hep-th/0002087.

\bibitem{sda}S.~P.~de Alwis,
``Brane world scenarios and the cosmological constant'',
hep-th/0002174.

\bibitem{Bremer}
M.~S.~Bremer, M.~J.~Duff, H.~L$\ddot{\rm u}$, C.~N.~Pope and K.~S.~Stelle,
``Instanton cosmology and domain walls from M-theory and string theory'',
Nucl.\ Phys.\  {\bf B543}, 321 (1999)
[hep-th/9807051].

\bibitem{ADS}N.~Arkani-Hamed, S.~Dimopoulos, N.~Kaloper and R.~Sundrum,
``A small cosmological constant from a large extra dimension'',
hep-th/0001197; 
\par S.~Kachru, M.~Schulz and E.~Silverstein,
``Self-tuning flat domain walls in 5d gravity and string theory'',
hep-th/0001206.

\bibitem{critique}
S. Gubser,
``Curvature singularities: The good, the bad and the naked'',
hep-th/0002160.

S. F${\ddot {\rm o}}$rste, Z. Lalak, 
S. Lavignac and H. P. Nilles,
``A comment on selftuning and vanishing cosmological constant in brane
world models'',
hep-th/0002164.

E. Witten,
``The cosmological constant from the viewpoint of string theory'',
hep-th/0002297.

J. Polchinski and M. J. Strassler,
``The string dual of a confining four-dimensional gauge theory'',
hep-th/0003136.

\bibitem{Berkovits}
N.~Berkovits, C.~Vafa and E.~Witten,
``Conformal field theory of AdS background with Ramond-Ramond flux'',
JHEP {\bf 9903}, 018 (1999) [hep-th/9902098]; 
\par ``Quantization of the superstring in Ramond-Ramond backgrounds'',
Class.\ Quant.\ Grav.\  {\bf 17}, 971 (2000)
[hep-th/9910251].

\bibitem{deW}
O.~DeWolfe, D.~Z.~Freedman, S.~S.~Gubser and A.~Karch,
``Modeling the fifth dimension with scalars and gravity'',
hep-th/9909134.

\bibitem{KL}
R.~Kallosh and A.~Linde,
``Supersymmetry and the brane world'',
JHEP {\bf 0002}, 005 (2000)
[hep-th/0001071].

\bibitem{RS1} 
L.~Randall and R.~Sundrum,
``A large mass hierarchy from a small extra dimension'',
Phys.\ Rev.\ Lett.\  {\bf 83}, 3370 (1999) [hep-ph/9905221]. 

\bibitem{RS2}
L.~Randall and R.~Sundrum,
``An alternative to compactification'',
Phys.\ Rev.\ Lett.\  {\bf 83}, 4690 (1999)
[hep-th/9906064].

\bibitem{Verlinde}
H.~Verlinde,
``Supersymmetry at Large Distance Scales'',
hep-th/0004003.

\bibitem{pw}
J. Polchinski and E. Witten,
``Evidence for heterotic-type $I$ duality'',
Nucl. Phys. {\bf B460} (1996), 525.

\bibitem{HW}P.~Horava and E.~Witten,
``Eleven-Dimensional Supergravity on a Manifold with Boundary,''
Nucl.\ Phys.\  {\bf B475}, 94 (1996)
[hep-th/9603142]; 

P.~Horava,
``Gluino condensation in strongly coupled heterotic string theory,''
Phys.\ Rev.\  {\bf D54}, 7561 (1996)
[hep-th/9608019].

\bibitem{M}
A. Lukas, B. A. Ovrut, K. S. Stelle and D. Waldram,
''Heterotic M-theory in five-dimensions'',
Nucl. Phys. {\bf B552} (1999) [hep-th/9806051].
 
C.~S.~Chan, P.~L.~Paul and H.~Verlinde,
``A note on warped string compactification,''
hep-th/0003236.

M. Spalinski and T. R. Taylor,
''Branes and fluxes in $D=5$ Calabi Yau compactifications of M theory'',
hep-th/0004095.

B. R. Greene, K. Scham and G. Shiu,
''Warped compactifications in M and F theory'',
hep-th/0004103.

\bibitem{vacuum}
A. Khavaev, K. Pilch and N. P. Warner,
``New vacua of ${\cal N}=8$ supergravity in five dimensions.''
hep-th/9812035.

\bibitem{ST} 
K. Skenderis and P. Townsend,
``Gravitational stability and renormalization group flow'',
Phys. Lett. {\bf B468} (1999), 46.

\bibitem{clp}
M. Cveti$\breve{\rm c}$, H. L$\ddot{\rm u}$ and C. N. Pope,
``Domain walls and massive gauged supergravity potentials'',
hep-th/0001002.

\bibitem{BKV}E.~Bergshoeff, R.~Kallosh and A.~Van Proeyen,
``Supersymmetry in singular spaces,''
hep-th/0007044.

\bibitem{DLS}M.~J.~Duff, J.~T.~Liu and K.~S.~Stelle,
``A supersymmetric type IIB Randall-Sundrum realization,''
hep-th/0007120.

\bibitem{Kraus}P.~Kraus,
``Dynamics of anti-de Sitter domain walls,''
JHEP {\bf 9912}, 011 (1999)
[hep-th/9910149].

\bibitem{squashed}
K. Pilch and N. P. Warner,
``A new supersymmetric compactification of chiral IIB supergravity'',
hep-th/0002192.

\end{thebibliography}
\end{document}